# Tropical and Extratropical Cyclone Detection Using Deep Learning

**Christina Kumler-Bonfanti**

University of Colorado Boulder

Cooperative Institute for Research in Environmental Sciences

NOAA/OAR/ Global Systems Laboratory

Boulder, CO

Christina.E.Kumler@noaa.gov

| **Jebb Stewart** | **David Hall** | **Mark Govett** |
|---|---|---|
| NOAA/OAR/ Global Systems Lab | NVIDIA Corporation Lafayette, CO | NOAA/OAR/ Global Systems Lab |
| 325 Broadway | dhall@nvidia.com | 325 Broadway |
| Boulder, CO | | Boulder, CO |
| Jebb.Q.Stewart@noaa.gov | | Mark.W.Govett@noaa.gov |

**Abstract:**

Extracting valuable information from large sets of diverse meteorological data is a time-intensive process. Machine learning methods can help improve both speed and accuracy of this process. Specifically, deep learning image segmentation models using the U-Net structure perform faster and can identify areas missed by more restrictive approaches, such as expert hand-labeling and a priori heuristic methods. This paper discusses four different state-of-the-art U-Net models designed for detection of tropical and extratropical cyclone Regions Of Interest (ROI) from two separate input sources: total precipitable water output from the Global Forecasting System (GFS) model and water vapor radiance images from the Geostationary Operational Environmental Satellite (GOES). These models are referred to as IBTrACS-GFS, Heuristic-GFS, IBTrACS-GOES, and Heuristic-GOES. All four U-Nets are fast information extraction tools and perform with a ROI detection accuracy ranging from 80% to 99%. These are additionally evaluated with the Dice and Tversky Intersection over Union (IoU) metrics, having Dice coefficient scores ranging from 0.51 to 0.76 and Tversky coefficients ranging from 0.56 to 0.74. The extratropical cyclone U-Net model performed 3 times faster than the comparable heuristic model used to detect the same ROI. The U-Nets were specifically selected for their capabilities in detecting cyclone ROI beyond the scope of the training labels. These machine learning models identified more ambiguous and active ROI missed by the heuristic model and hand-labeling methods commonly used in generating real-time weather alerts, having a potentially direct impact on public safety.

# 1. Introduction:

Enormous quantities of data are being generated by increasingly higher-resolution models, next-generation satellites, radar, and a myriad of ground-based in situ observation data. Only a small percentage of these data are processed and utilized in real-time applications, such as severe weather alerts, numerical weather prediction (NWP) models, and data assimilation (DA) due to the time required to process it. In addition, detection of severe weather in model output or satellite images relies upon trained experts to identify significant features such as cyclones or weather fronts. However, this process takes time and is subject to human bias and error.

Historically, heuristic rule-based models have been used to automatically identify strong storms by using a set of well-defined rules. However, complex weather phenomena are difficult to quantify and heuristic models tend to be fragile. The Community Effort to Intercompare Extratropical Cyclone Detection and Tracking Algorithms (IMILAST) compiles an analysis of extratropical cyclone trackers that both qualitatively and quantitatively compare the performance of these types of models, detailing the limitations of each. For example, some models are limited to the ocean and do not function over land [Neu et al. 2013]. These methods can be accurate when the feature of interest is well defined, but struggle to identify more ambiguous regions. Tropical cyclones are well defined as any closed, low-level rotating system with thunderstorms and high winds that originate over the waters of the tropics [Holland 1993 and Merrill 1993]. However, there is no universally agreed-upon definition for extratropical cyclones. NOAA defines extratropical cyclones as low-pressure systems that could be associated with a front [AOML] while the American Meteorological Society (AMS) defines them as any cyclonic-storm that is not a tropical-cyclone [AMS]. Heuristic-based models typically require a specific set of meteorological variables provided by weather model outputs and often cannot be run directly with observations from sources such as satellites without further information or routines.

Machine Learning (ML) techniques can be used effectively to identify a range of different meteorological features that may be broadly classified as Regions Of Interest (ROI). For instance, ML methods have successfully been used to classify certain types of thunderstorms [Jergensen 2020]. In contrast with explicit heuristic approaches, ML techniques learn to define features implicitly by working backward from a given set of examples. Neural Networks (NNs) are a type of ML techniques which take inspiration from human cognition and are able to perform many tasks previously beyond the reach of explicit rule-based approaches. An advantage of this approach is that NNs can more easily identify complex and ambiguous phenomena that are not easily described by simple heuristics. Additionally, a trained NN is often much faster than a hand-coded heuristic model making it more practical for real-time identification.

There are three main reasons NNs have become increasingly popular for image recognition: increasing quantities of data, the exponential growth of computational resources, and the ready availability of modern ML tools, such as Python libraries, that simplify the process of model development. NNs have been trained to identify or classify both clear or ambiguous objects on large datasets, such as Modified National Institute of Standards and Technology database (MNIST), which classifies written digits as numbers from zero to nine, and ImageNet, which classifies objects contained in a wide variety of images [Deng 2012 and Deng 2009]. In some cases, NNs have been trained to perform object recognition tasks at superhuman levels. Such comparisons demonstrate enormous potential benefits in using NNs for classification in many earth science fields [Ciresan 2012].

Convolutional neural networks (CNNs) are a class of NNs that have proven to be particularly effective for image analysis. They rely upon one or more convolutional layers that are able to recognize important features regardless of where that feature appears within the image. These NNs are particularly effective for image problems because they use small segments or squares of input data and retain spatial information by surrounding pixels needed for analysis. CNNs are used in a variety of science applications including medical scan analysis, signal processing, short-term prediction, image recognition, identification, and segmentation tasks [Cheng et al. 1996, Dia 2001, Cichoki and Unbehauen 1996, Lawrence et al. 1997, Liang and Hu 2015, and Giacinto and Roli 2001]. AlexNet and VGGNet, two of the most popular and widely used CNNs, were trained with millions of images and are highly accurate [Krizhevsky 2017 and Simonyan et al. 2015].

Deep Learning (DL) is a term used to describe the optimization of NNs with multiple layers. Deep networks tend to be large relative to other ML models and their training typically requires powerful GPU acceleration in order to make training practical. In comparison with other approaches, deep CNNs are particularly effective for image classification and segmentation [LeCun 2015, He et al. 2016, and Liu and Deng 2015]. DL models are increasingly being applied to complex problems in earth science and meteorology, including probabilistic hail forecasting, cloud classification, predicting algal blooms, tropical cyclone track forecasts, and severe weather detection [Gagne et al. 2017, Giffard-Roisin 2020, Lee et al. 2004, Recknagel 1997, and Mcgovern 2017].

This paper describes work using DL to perform image segmentation for the detection of tropical and extratropical cyclone ROI. The automated detection of significant ROIs is potentially valuable for improving data assimilation and model initialization for numerical weather prediction models, amongst many other possible applications. For example, Figure 1 shows a water vapor image produced by an older generation NOAA Geostationary Operational Environmental Satellites (GOES). In this figure, red boxes indicate ROIs for hurricane Harvey and other cyclones. DL segmentation enables the rapid, automatic identification of specific regions of interest from high volume datasets such as GOES satellite observations. This type of automation is critical due to the enormous volume of data waiting to be processed. In particular, NOAA's latest generation GOES 16 and GOES 17 have 16 bands that take data samples every half second to five minutes at ½ km to 2 km resolution [Schmit 2018]. Only an estimated 3-7% of observational data are selected to be used in NWP models, and an even smaller fraction of that data are actually assimilated into the models [Weingroff 2014]. GOES-16 and GOES-17 produce over 100 times as much data as the previous GOES missions, with high potential value for NWP. With the current system, the amount of data far exceeds the computing time available to process it and instead, simple data thinning techniques are applied and the majority of the data is discarded. In contrast, targeted selection of data guided by DL can intelligently select highly relevant observations, providing additional data in regions that are active or soon to be active.

There are a number of DL models that may be used to solve problems in the earth sciences. The selection of the right model depends on the type of problem being solved. For image segmentation tasks, the U-Net deep convolutional neural network has become a de-facto standard choice. The U-Net structure is specifically designed for image segmentation tasks. It employs an encoder-decoder structure with skip connections that is able to identify important features at multiple spatial scales. Although originally designed for medical segmentation tasks, the U-Net has proven successful at extracting both the coarse and fine-scale features important for determining events of interest in meteorological data [Ronneberger 2015 and 2017]. Deep convolutional networks, such as U-Net, tend to identify simple features in the first few layers, which are combined to form high-level features in subsequent layers [LeCun and Bengio 1995]. Both low-level and high-level features are learned simultaneously in an iterative fashion as the model learns to fit the training data.

The paper is organized as follows. In Section 2, the U-Net architecture is described, as are the numerical metrics used to evaluate its success. Section 3 describes both qualitatively and quantitatively the design and performance of the best U-Net model obtained for identifying tropical cyclones using the Global Forecast System (GFS) total precipitable water field as inputs. In Section 4, three additional U-Net models are introduced that identify both tropical and extratropical cyclones from GFS total precipitable water output data as well as GOES water vapor data inputs. The last section provides a summary of the U-Net approach and offers further discussion on potential applications in NWP.

**2. Deep Learning Architecture: The U-Net**

The following four subsections describe the programming environment, the model design, performance metrics, and the CPU and GPU Performance measurements.

2.1) Programming Environment

The U-Net models were written with the Keras NN Application Programming Interface (API) in Python because of its user-friendly interface [Gulli et al. 2017]. Keras is simple to use and well suited both for model prototyping and deployment. Keras offers several framework options "under the hood" and Google's TensorFlow framework was selected for its ease of use and its performance on multiple GPU machines. This choice of framework enables training, analysis, and visualization to be readily combined in a single script.

2.2) U-Net Design

The cyclone ROI U-Net models have several dynamic tuning variables common to DL models including but not limited to: number of convolution layers, pooling layers, rectifier activation functions, and loss functions. Unique to the U-Net structure, there are additional up-convolutions and skip-connections, which connect relevant information that might be lost in the pooling layer to the expanding path. The structure depth varies depending on the number of convolution and pooling layers, where each pooling layer reduces the size of the feature map. The cyclone ROI U-Net models each had unique depths for optimal model performance; Figure 2 illustrates the general U-Net structure that was used for all U-Net models. As the U-Net contracts, or encodes, it extracts generalized information from the input image. During the U-Net expansion, or decoder part, skip-connections join generalized information with more localized information, often represented graphically as a "U" shape. Each cyclone ROI U-Net model has batch normalization after the convolution, which normalizes the inputs into the next layer, to increase training performance and time.

Additional parameters that were tuned for the cyclone ROI U-Net models were the activation function, dropout or noise, and loss functions. The activation function is used at each convolution and after tuning experiments the rectified linear unit (ReLu) was the best performing across the designed U-NETs. ReLu will take the convolution field as inputs and if the value is positive, keep it. If the value is negative, it sets it equal to zero [Glorot 2011]. The model can avoid overfitting with either noise or dropout additions to the U-Net. Adding "noise" is when random values are taken from something such as a gaussian curve and applied to the output at the end of a layer. Dropout is where randomly selected neuron units are turned off during the gradient process, essentially preventing the model from learning from certain neurons for that iteration [Srivastava 2014]. A loss function is how a DL model learns and different loss functions perform better depending on the type of dataset distribution.

The datasets used to train the U-Net models are divided into training, validation and testing groups. The training and validation data are used to build the model with its weights and features. The test data are kept hidden from the training and validating stages and treated as new, unseen inputs that evaluate the robustness of a model on completely separate input data. To avoid memorization and overfitting by the U-Net, the training, validation, and test data must be chosen carefully. For instance, the model can identify subsequent timesteps from long-lasting cyclones that persist through three or more image inputs and will overfit the model due to memorization. The cyclone ROI U-Net models accounted for this issue by having training, validation, and testing datasets that were separated by yearly chunks. The yearly breakup was determined based on available input data and labels.

Another parameter that contributes to either overfitting or poor performing models comes from selecting the batch and epoch sizes. A batch is a random subgroup of data selected from training data that is used in the training and validation stage. A single completed epoch is when all of the data in the training set have been in a batch and through that process. The training process works through more epochs, each refining the previously designed weights of the model. Selecting an optimized combination of the batch size of the inputs into the U-Net as well as the number of epoch iterations that are allowed to run produces the U-Net that converges on an optimized model. A model is determined to converge on the best model when the loss function, or training errors, is minimized and the training accuracy is high.

U-Net models can take as many input channels of data as memory allows so long as these inputs are all of the same dimension. To provide additional information to the cyclone U-Net models, all the cyclone ROI U-Net models were run with three input channels that were selected from the same data input field. The three channels represented data at the current t, t-1, and t-2 time steps for the starting t timestep. This method gave the U-Nets a sense of time, cyclone rotation, and performance improvements.

A segmentation DL model means that the cyclone ROI U-Net models have input image pixel values defined to be either 0 for no cyclone or 1 for yes cyclone. The model produced image pixel values that ranged from 0 to 1, where 1 is 100% likely or confident that the pixel is a cyclone pixel and 0% means no cyclone. There is one last parameter that was changed based on the size of the cyclone that was being segmented. Each cyclone center, represented by the labeled data latitude and longitude center point, is additionally labeled by a pixel bounding box that contains all 1's. This is done to offset any errors in cyclone center value as well as to encompass a larger area of the storm. It also provides more "yes cyclone" pixels in an image that contains mostly "no cyclone" pixels.

2.3) Evaluating U-Net Accuracy and Performance

The cyclone ROI problem is an imbalanced dataset of "no cyclone" to "yes cyclone" pixel labels where most pixels belong to the "no cyclone" category, where presenting results only in terms of accuracy is not sufficient. High accuracy for these cyclone U-Nets indicates that while they are good at detecting the non-cyclone events, it is unclear how good the models are for the yes-cyclone event detection. To understand the similarity between the truth labels and the U-Net model identified ROI, the coefficient values from the loss functions were evaluated, where the values range from [0, 1] and 1 represents a perfect match. The best ML models are designed to minimize the loss function. Four loss functions known to work better for imbalanced datasets were chosen: Binary Cross Entropy (BCE), Dice, Tversky, and focal loss.

i) The BCE loss function is mathematically defined as:

$$BCE(p, y) = -\log(p) \quad \text{if } y = 1$$

$$BCE(p, y) = -\log(1 - p) \quad \text{else}$$

Where this $p$ is the model probability between [0,1] and y is the classification value from the truth labeled point [Lin 2017]. This loss function is log-based in nature and penalizes instances where the model predicts the wrong result.

ii) The Dice and Tversky loss functions are Intersection over Union (IoU) methods that overlap regions of yes cyclone and no cyclone [Sudre 2017]. The Tversky coefficient assigns a weight to false positive ($\alpha$) and false negative ($\beta$) results, defined by:

$$TC = \frac{|X \cap Y|}{|X \cap Y| + \alpha |X - Y| + \beta |Y - X|}$$

Where the models defined $\alpha = 0.3$ and $\beta = 0.7$ to add emphasis on the yes cyclone instances. The Dice coefficient is defined similarly except $\alpha = \beta = 0.5$ so the function reduces to:

$$DC = \frac{2 * |X \cap Y|}{|X| + |Y|}$$

iii) The focal loss function is designed to reduce the weight of the predominant no-cyclone pixels so the CNN focuses on the harder and fewer, yes-cyclone pixels. For a two-class problem such as the cyclone ROI, focal loss is defined as:

$$FL(p) = -(1-p)^\gamma \log(p)$$

Notice that if γ = 0, the focal loss equals the cross-entropy loss function. In setting γ > 0, the loss of a nearly correctly labeled point becomes orders of magnitude less than if γ = 0. For the cyclone ROI models, γ = 2. This is a standard number for most implementations of this loss function.

2.4) CPU and GPU Performance

      The training period of U-Net and DL models can take long periods of time on Central Processing Units (CPUs). Training time improves substantially when the U-Net model uses Graphics Processing Units (GPUs) [Bahrampour et al. 2015]. Table 1 shows the difference between model training times from an experimental cyclone ROI U-Net model on a CPU and NOAA's GPUs. The dramatic difference in training times proved the need for GPU use in order to train models in a practical time frame. Both the data processing for U-Net inputs and U-Net model training processes were run on NOAA's GPU supercomputer system. The Horovod DL software-enabled running on multiple GPUs simultaneously and further reduced the U-Net training time. Horovod was designed by Uber to support both startup and speed up of DL using TensorFlow distributed on large scale systems [Sergeev 2018]. Once a model has been trained, the inference stage, where a trained model identifies ROI from unseen data input, is completed in a fraction of time.

3. Tropical Cyclone ROI U-Net Model Results

      The U-Net is a type of supervised DL model which performs best when trained with substantial quantities of broadly representative, labeled data. The labeled data provides the

"truth" labels for the input training data, indicating which areas of an input image encompass a cyclone. This section will discuss the tropical cyclone ROI U-Net model, which will be referred to as the IBTrACS-GFS U-Net. It was trained using truth labels from the International Best Track Archive for Climate Stewardship (IBTrACS) tropical cyclone database and detected the locations of the tropical cyclones in the GFS total precipitable water output data field [Knapp et al 2010]. To remain consistent with the Simpson Saffir scale, a 34 knots or higher wind threshold was applied to the IBTrACS database [National Hurricane Center 2013]. Since the GFS total precipitable water output data field is one-half degree resolution in both latitude and longitude, there is noticeable data skewing in projection near the poles. GFS data are blended short term forecasts with the latest observations, through a DA process, and are produced every six hours. Each file contains the best approximation of the state of the atmosphere at the zero hour time. Additional GFS forecasts are provided within the same file. The IBTrACS-GFS U-Net used input data from both the zero-hour (analysis) and three-hour forecast within each GFS file. The three-hour forecast was treated like the following time-step's zero-hour analysis to provide more input training data. Therefore, there were total precipitable water outputs available at 0, 3, 6, 9, 12, 15, 18, and 21 UTC per day. There is a small difference in the total precipitable water field's forecast and actual observed state, resulting in a cyclone center label that might not be over the correct spot in the image. However, this difference in location is acceptable because it still remains within the bounded region for a "yes" cyclone. Using the three-hour GFS forecast is appropriate for slower-moving, longer living cyclone systems because the additional difference in storm center location is also within the bounding region.

Row 1 in Table 2 shows the combination of tuning parameters that produced the best-performing IBTrACS-GFS U-Net model (the subsequent rows show the best combinations of tuning parameters for U-Nets that used other combinations of labelled data and input data). The pixel bounding region was 25x25 pixels from the GFS (approximately 300km square) and the input image size was the full GFS resolution of 720x361. The Tversky loss was the best performing loss function for this U-Net and the model structure was six-block layers deep. There were 8,622 training samples over the four-year period and it took the U-Net roughly 36 minutes to train on GPUs over the course of 37 epochs. Once the IBTrACS-GFS U-Net was trained, it quickly identified tropical cyclone ROI from a test image in roughly 0.03 seconds. Table 3 shows the statistical testing performance comparing the truth, based on the labeled inputs, to the U-Net produced ROI. The IBTrACS-GFS U-Net, row two, achieved an accuracy of 99% and high Dice and Tversky coefficients of 0.75 to 0.76 (Table 3, row 1).

In Figure 3, there are four images that progress forward three hours in time each panel. Within each panel, the truth labeled image is above the IBTrACS-GFS U-Net model ROI labeled image. The IBTrACS-GFS U-Net tended to identify regions as boxes and this is likely due to the square areas of interest that trained the model. The model captures nearly all tropical cyclones within all ocean basins and has no issues with detection near the edges of the domain. Since these cyclones are circular, very bright (wet) in their signature, and fairly uniform in appearance, the model rarely identifies a region that doesn't appear somewhat as a tropical cyclone. Both well-developed tropical cyclones with identifiable eye features as well as weaker or newly forming/dissipating tropical cyclones are detected by the U-Net. The U-Net does not always detect ROI consistently in time. There are instances when a ROI would go away prematurely for a time step and then might come back and continue detection. Other times, the truth label terminated prematurely, which might explain these instabilities in how the model learned the behavior. One hypothesis for this behavior is if the IBTrACS labels start late or terminate early because of the applied wind threshold, then the U-Net has shown to learn weaker tropical cyclone structure and can continue to track the storm. This is because the U-Net was trained without the wind threshold knowledge explicitly and likely learns more from cyclone structure. Another hypothesis is that if a tropical cyclone weakens and its structure breaks up, then it might still have the wind speed required to keep the truth label but be more ambiguous of an object for the U-Net to detect at each time step. With these additional tropical cyclone ROI labels, this is beneficial for early detection of storms that might become tropical cyclones or regions of cyclonic potential.

Figure 3 is plotted to show the total likelihood of cyclone regions. Lower confidence areas are those with a numerical value closer to zero and are indicated by the lighter blue-colored boxes. Higher confidence values are those closer to one and are shown in more red-colored boxes. Occasionally, very bright signatures in the total precipitable water field that are non-tropical cyclone regions along the Intertropical Convergence Zone (ITCZ) are labeled as ROI by the U-Net. While rarely these ROI instances had a high confidence (red), the majority of these cases were labeled with a lower confidence (blue) and could be eliminated if a threshold were applied to the U-Net ROI labels. This figure shows the U-Net model tropical cyclone ROI that are labeled outside of the tropics. These regions have a lower threshold confidence but since the model was not trained to be constrained in latitude, it has the potential to track tropical cyclones as through their transition as they become extratropical.

## 4. Variances of U-Net Model for Extratropical Cyclone Detection

Based on the success of the U-Net model in section 3, the same approach to investigate the DL detection of tropical cyclones was extended to the detection of extratropical cyclones with either GFS total precipitable water data as inputs or GOES water vapor satellite data as inputs. The extratropical cyclone data labels were collected from prior work by Bonfanti et. al. [2018]. The three new U-Net variant models created will be referred to as Heuristics-GFS U-Net, IBTrACS-GOES U-Net, and Heuristics-GOES U-Net. Additionally, other fine-tuning components in the U-Net, such as model depth or the loss functions, were modified to optimize each individual U-Net's performance. The following three subsections describe each U-Net model.

4.1) Heuristic Cyclone Labels with GFS Inputs

Defining extratropical cyclones is a harder task than identifying tropical cyclones due to the diversity in water vapor signatures and varying definitions for extratropical cyclones. To avoid a time-consuming and subjective process of hand labeling extratropical cyclones, the table outputs from the heuristic cyclone labeler by Bonfanti et al [2018] was used to provide the truth labels. That heuristic model was designed for machine learning applications and did not label cyclones in polar regions. The work identified storms over land masses, providing the U-Net model with more training examples than other heuristic model options. However, a heuristic model introduces its own set of biases and missed cyclones and heuristics datasets are not fully inclusive of all events. Given the nature of ambiguity in the definition of extratropical cyclones, there are discrepancies between truth labeled datasets on correctly labeled cyclone storms. This made it difficult to evaluate the numerical performance of the U-Net. This is not directly quantifiable and therefore difficult to extract from numerical results, but differences between truth and U-Net labeled ROI will lower the coefficients and accuracy. For instance in training, the model might learn and therefore identify an event as a "no" when it could be a "yes" due to or vise-versa. These issues are addressed in the qualitative analysis. Given those challenges, the U-Net trained from these labels still identify ROI that are interesting, valuable, and potentially correct even if the truth label is missing. The goal of this research was to identify regions of cyclonic interest, and therefore regions that are missed in a truth label but identified in the U-Net are of good value.

      The Heuristic-GFS U-Net performed slightly poorer than the IBTrACS-GFS U-Net model because of the inclusion of the extratropical cyclones. Table 2 shows that both the loss function and depth that produced the optimal Heuristic-GFS U-Net were different than the IBTrACS-GFS U-Net, having a depth of only 5 and using the Dice loss. The shallower the U-Net, the fewer hyperparameters the model has to tune. One potential reason that the Heuristic-GFS U-Net has a shallower optimized U-Net is that it has a larger training size. This may mean that fewer hyperparameters are needed to converge on the better performing model. One of the biggest differences between the IBTrACS-GFS U-Net and this Heuristic-GFS U-Net was the bounding box size, which was increased from 25x25 to 30x30. Since extratropical cyclones are bigger than tropical cyclones, increasing the size from 25 to 30 improved the U-Net performance by encompassing the larger areas of extratropical cyclones. Row 2 of Table 3 shows that the accuracy of this model remains high, at 80%, but the Dice and Tversky coefficients were lower than the IBTrACS-GFS U-Net at 0.5 and 0.6 respectively.

      This U-Net model has a fast inference stage on GPU, identifying extratropical and tropical cyclone ROI from an input in 0.03 seconds. It identified these ROI much faster than the heuristic model used to create the data labels on the same input source. The inference time for the heuristic model (used to create the labels) for a month's worth of cyclone ROI took 18.67

seconds while the Heuristic-GFS U-Net ran in 6.48 seconds. This shows that the U-Net model identified cyclone ROI 2.88 times faster than the heuristic method.

The interest in identifying extratropical cyclone ROI extends beyond matching truth labels to U-Net labels because there is much value, such as early detection, in the regions uniquely labeled by the U-Net. To better understand the behavior of which types of cyclones were identified as ROI in the Heuristic-GFS U-Net, a qualitative analysis was completed. This was done on images like Figure 4 and compared the labeled truth labeled GFS data (top) with the U-Net identified cyclone ROI (bottom). Figure 4 has a threshold applied to the U-Net ROI of 70% confidence, meaning that ROI are only colored if there is a chance of at least 70% cyclone likely. The figure shows areas of missed ROI detection as well as an area of false ROI detection. An explanation for this behavior is that GFS total precipitable water output data alone might not provide enough information to the U-Net in the high latitude regions to distinguish between different types of extratropical cyclones as well as non-cyclone events. A different reason is that the extratropical cyclones heuristic truth labels might have incorrectly missed a yes cyclone label for certain cyclones due to the heuristic set of rules. This would incorrectly train the U-Net to miss extratropical cyclones by assigning an error score to what should have been a correctly identified cyclone but that the truth label said was not a cyclone. This case is seen in Figure 4 with the missed identification of a tropical cyclone in the heuristic truth label but where the U-Net had labeled it as an ROI. This makes quantitative analysis of U-Net performance alone incomplete. The qualitative analysis proves that the U-Net provided an important label on ROI that was missed by a different method.

Based on the analysis concluded by IMILAST, there is a climatological pattern of more frequent labels in the southern hemisphere than the northern hemisphere. The Southern Ocean has more extratropical cyclones and the Heuristic-GFS U-Net identifies the more ambiguous cyclone events that occur in that ocean basin because of that learned trait. Since there is no distinction between tropical and extratropical cyclones, this model also will track cyclones from tropics through the extratropics.

4.2) IBTrACS Tropical Cyclones Labels with GOES Inputs

This U-Net model uses the IBTrACS labels with GOES water vapor imagery. The most recent GOES satellite imagery ranges from 75 degrees west to 135 degrees west longitude, so unlike the previous two U-Net models, this does not cover the whole globe [Jenner 2017]. Only the water vapor channel (6.48um), with a resolution of 4 km to 8 km, was used as input. Tropical cyclone signatures are detectable by the trained human eye in water vapor from a single time-step or in a series of sequential images as distinct, bright, small circles. If the cyclone is well-developed, it sometimes has the prominent "eye" feature. That signature persists at any time or lighting condition during an earth day, making this IBTrACS-GOES U-Net model useful in identifying potential or current tropical cyclones in real-time applications.

Table 2 row 3 shows there are 5,638 test samples for the IBTrACS-GOES U-Net and that the U-Net is five layers deep. The BCE loss function gave the best performing model for these inputs and labels. Since GOES satellite imagery is very large and with the complexity of the U-Net model structure, the data had to be resized to 1024X512 to fit in GPU memory. Similar to the IBTrACS-GFS U-Net, the IBTrACS-GOES U-Net had a high accuracy of 99%. Table 3 shows a lower Dice and Tversky coefficient of about 0.7, indicating a slightly poorer ROI detection performance for GOES data inputs than GFS.

This U-Net might have had a harder time identifying ROI because of scanning gaps or projection skewing in the satellite imagery because the angle of observation of the earth is more oblique as it approaches the outer edges of the imagery. There were small horizontal data gaps in some of the data inputs due to satellite scanning issues. Figure 5 indicates the potential for errors near the boundaries of the satellite images due to curvature and skewing of the data. Despite skewing, the IBTrACS-GOES U-Net performed generally well, but when it missed ROI, it missed them near the boundaries more than for storms centered in the image. Finally, the greatest impact on the U-Net results may have been the smaller sample size available for training. The IBTrACS-GOES U-Net had three thousand fewer training samples largely due to the smaller coverage area.

Relying on qualitative metrics alone is not sufficient. The broader definition of cyclones in the data suggests early detection of cyclones that might not yet be identified in the IBTrACS database or with heuristic truth labels. Identification of such cases would be beneficial for real-time alert systems but are scored negatively in the quantitative estimates. For example, not shown in the figure are instances when the U-Net incorrectly identified a non-cyclone weak tropical storm or area of tropical convection in the ITCZ. This is accurately scored in the numerical quantitative results. However, Figure 5 (plotted with a 70% yes-cyclone pixel confidence threshold) indicates that there are instances of early detection as well as false positives. The U-net is said to provide early detection when the ROI segmentation appears in images in time over a region that the truth label later identifies. Panels 1, 2, and 3 show that the U-net identified a ROI before the truth label appeared in the final panel 4. The IBTrACS-GOES U-Net correctly identifies the two truth-labeled ROI cyclones. It further detects an additional region in the tropics that has a very similar signature to a hurricane that persists through all four timeframes. The U-Net additionally identifies a cyclone ROI transitioning into the extratropics.

4.3) Heuristic Cyclone Labels with GOES Inputs

The Heuristic-GOES U-Net used the heuristic labels from Bonfanti et.al. [2018] and inputs from the GOES water vapor data. This U-Net had the most training samples with 25,288 and the largest bounding box (60 pixels). This larger bounding box is the biggest difference between this U-Net model and the other models listed in Table 2. Extratropical cyclones consume more pixel area and therefore warrant a larger bounding size. The best configuration of the model was the shallowest U-Net, with four layers giving the optimal model, and used the Tversky loss function. One guess as to why the shallower U-Net model gave the optimized model for these labels and input source is because it generalized better to differences in satellite extratropical cyclone appearance. The Heuristic-GOES U-Net also had an ambiguous interpretation of both quantitative and qualitative results similar to the Heuristic-GFS U-Net due to both a broader definition of cyclones and early detection of cyclones in the GOES input data. The quantitative results can be misleading. Table 3 shows that the Heuristic-GOES U-Net had a higher accuracy at 90% than its GFS counterpart, but that it had the lowest Dice and Tversky coefficient values at about 0.5. While that value is lower than the other models, visually there is still much agreement between the truth labels and Heuristic-GOES U-Net identified ROI.

Similar to the Heuristic-GFS U-Net, the Heuristic-GOES U-Net had a pattern of more frequent cyclone ROI detections in the southern hemisphere than the northern hemisphere. Figure 6 compares truth labels from the heuristic model to the Heuristic-GOES U-Net detected ROI. The plots have a threshold confidence of at least 70% chance that the pixel contains a cyclone. The U-Net had a tendency to correctly label fewer ROI along the boundary edges than in other regions in the satellite image and had occasional noise in the detection along the boundaries. There was a higher quantity of Heuristic-GOES U-Net detected ROI than truth labels in general. This figure was selected to show an example of when the Heuristic-GOES U-Net correctly labeled all three tropical cyclones while the truth labels had not yet identified them all. False detection by the U-Net occurred when it incorrectly labeled a very small ROI in the ITCZ as a cyclone when they are not. Similar cases were observed in the other U-Nets and it has been discussed how this behavior degrades the quantitative performance of the model. It was more difficult to qualitatively analyze extratropical cyclone U-Net detected ROI due to the diversity in extratropical cyclone appearance in the water vapor data.

Given the variety of types and appearances in the water vapor channel of extratropical cyclones, the Heuristic-GOES U-Net model impressively identified diverse looking patterns for extratropical cyclones at the same time as detecting brighter, smaller tropical cyclones. The most common extratropical ROIs detected in the U-Net have clear comma-shaped wet-dry signatures, indicating rotation. This is expected since it is also an easier signature for humans to identify as well and indicates a well-developed cyclone. In general, there were more false-labels and noise in the Heuristic-GOES U-Net than in both IBTrACS U-Nets, but as is the case in all other models, almost all non-cyclone events for the Heuristic-GOES U-Net remained correctly unlabeled. This indicates that the model can be used for fast, early cyclone detection as well as detection of ROI.

5. **Summary and Discussion**:

Four individual U-Net models were created to detect cyclone ROI from two different data sources and two different labeling sources: IBTrACS-GFS, Heuristic-GFS, IBTrACS-GOES, and Heuristic-GOES. NOAA's multi GPU system was used to significantly decrease the data processing and U-Net model training times. Once trained, the inference time for the Heuristic-GFS U-Net ran 3 times faster than the heuristic model used to create the labels from the same GFS input. This comparison shows that deep learning models can extract cyclone location information from GFS data faster than heuristic methods can. All of the models achieved a relatively high level of accuracy, ranging from 81% (Heuristic-GFS) to 99% (both IBTrACS U-Nets). IoU metrics were also used for evaluation. The Tversky coefficients for the Heuristic-GOES U-Net, Heuristic-IBTrACS U-Net, IBTrACS-GOES U-Net, and IBTrACS-GFS U-Net models respectively were: 0.558, 0.649, 0.680, and 0.750. These results show the U-Nets were optimized without overfitting and gave good results for diverse cyclone event detection. Further improvements to the U-Nets that could be explored in the future include newer activation functions, types of convolutions, increased data input.

The performance of these U-Net models proves that for image segmentation tasks in meteorology and climatology related fields, ML and DL models provide unique and faster alternatives to existing methods. Aside from numerical metrics, the U-Nets identified unique ROI that were not included within the truth labeled dataset. The U-Net models have versatile applications, such as cyclone ROI extraction in large datasets of high resolution when it is expensive to hand label extreme events of interest or impractical to identify ROI in real-time applications. These models show that when compared to the truth labels, they identify a high number of the same ROI area as heuristic or hand-labeling methods, but also uniquely identify additional regions that are missed by more traditional methods. This addresses the issue that labeled datasets are not always inclusive of all events and will impact how the U-Nets are evaluated numerically. It shows the benefit of integrating DL methods for diversifying an event database and for real-time applications, such as actively identifying ROI for potential cyclogenesis. Likewise, U-Nets provide fast detection of regions with high-impact weather in a time-sensitive scenario. They label the location information for areas with active or high-potential weather that benefits numerous weather products, such as aviation warnings and DA. Figure 7 highlights the tropical cyclone information in GOES water vapor data extracted from the IBTrACS-GOES U-Net as an example of what can be output at sub-second speeds to quickly locate ROI. Such information provides more data points for better understanding and forecasting the cyclone system.

The U-Net architecture used in these models is extensible to training new models for the detection and classification of other types of weather events. U-Nets can identify patterns of non-cyclone ROI, such as convection or convection initiation. The application of the U-Net can produce real-time information on active or high-potential weather locational information. It can scale to climatological settings by providing information on past weather events in climate data. These types of DL models are under-utilized. They show increasing success for ROI detection and their performance encourages the development of more DL models for future ROI detection schemes in a multitude of weather and climate applications.

**Acknowledgments**:

A special thank you to our colleagues, especially Isidora Jankov, who have provided help, feedback, and support through the duration of this project. An additional thank you to our machine learning colleagues in the Boulder area who have provided exposure to new methods and ideas. Christina Kumler is supported by funding from NOAA Award Number NA17OAR4320101. Jebb Stewart was supported by funding from NOAA Award Number NA14OAR4320125. This work was supported by NOAA's Software Engineering for Novel Architectures (SENA) program through the NOAA Office of the Chief Information Officer.

Deng, J., Dong, W., Socher, R., Li, L.-J., Li, K., & Fei-Fei, L. (2009). ImageNet: A large-scale hierarchical image database. *2009 IEEE Conference on Computer Vision and Pattern Recognition*. doi:10.1109/cvpr.2009.5206848

Deng, L. (2012). The MNIST Database of Handwritten Digit Images for Machine Learning Research [Best of the Web]. *IEEE Signal Processing Magazine*, *29*(6), 141–142. doi:10.1109/msp.2012.2211477

Dia, H. (2001). An object-oriented neural network approach to short-term traffic forecasting. *European Journal of Operational Research*, *131*(2), 253–261. doi:10.1016/s0377-2217(00)00125-9

Extratropical cyclone. *Extratropical cyclone - AMS Glossary*. http://glossary.ametsoc.org/wiki/Extratropical_cyclone. Accessed 30 October 2019

Gagne, D. J., Mcgovern, A., Haupt, S. E., Sobash, R. A., Williams, J. K., & Xue, M. (2017). Storm-Based Probabilistic Hail Forecasting with Machine Learning Applied to Convection-Allowing Ensembles. *Weather and Forecasting*, *32*(5), 1819–1840. doi:10.1175/waf-d-17-0010.1

Giacinto, G., & Roli, F. (2001). Design of effective neural network ensembles for image classification purposes. *Image and Vision Computing*, *19*(9-10), 699–707. doi:10.1016/s0262-8856(01)00045-227

Jenner, L. (2015, March 10). GOES Overview and History. *NASA*. NASA. http://www.nasa.gov/content/goes-overview/index.html. Accessed 25 September 2018

Jergensen, G. E., Mcgovern, A., Lagerquist, R., & Smith, T. (2020). Classifying Convective Storms Using Machine Learning. *Weather and Forecasting*, *35*(2), 537–559. doi:10.1175/waf-d-19-0170.1

Knapp, K. R., Kruk, M. C., Levinson, D. H., Diamond, H. J., & Neumann, C. J. (2010). The International Best Track Archive for Climate Stewardship (IBTrACS). *Bulletin of the American Meteorological Society*, *91*(3), 363–376. doi:10.1175/2009bams2755.1

Krizhevsky, A., Sutskever, I., & Hinton, G. E. (2017). ImageNet classification with deep convolutional neural networks. *Communications of the ACM*, *60*(6), 84–90. doi:10.1145/3065386

Lawrence, S., Giles, C., Tsoi, A. C., & Back, A. (1997). Face recognition: a convolutional neural-network approach. *IEEE Transactions on Neural Networks*, *8*(1), 98–113. doi:10.1109/72.554195

LeCun, Y., & Bengio, Y. (1995). Convolutional networks for images, speech, and time series. *The handbook of brain theory and neural networks*, *3361*(10), 1995.
29

Srivastava, N., Hinton, G., Krizhevsky, A., Sutskever, I., & Salakhutdinov, R. (2014). Dropout: a simple way to prevent neural networks from overfitting. *The journal of machine learning research*, *15*(1), 1929-1958.

Sudre, C. H., Li, W., Vercauteren, T., Ourselin, S., & Cardoso, M. J. (2017). Generalised Dice Overlap as a Deep Learning Loss Function for Highly Unbalanced Segmentations. *Deep Learning in Medical Image Analysis and Multimodal Learning for Clinical Decision Support Lecture Notes in Computer Science*, 240–248. doi:10.1007/978-3-319-67558-9_28

Weingroff, M. *How Satellite Observations Impact NWP*.
http://kejian1.cmatc.cn/vod/comet/nwp/sat_nwp/print.php.htm. Accessed 15 March 2019

**Figures and Captions:**

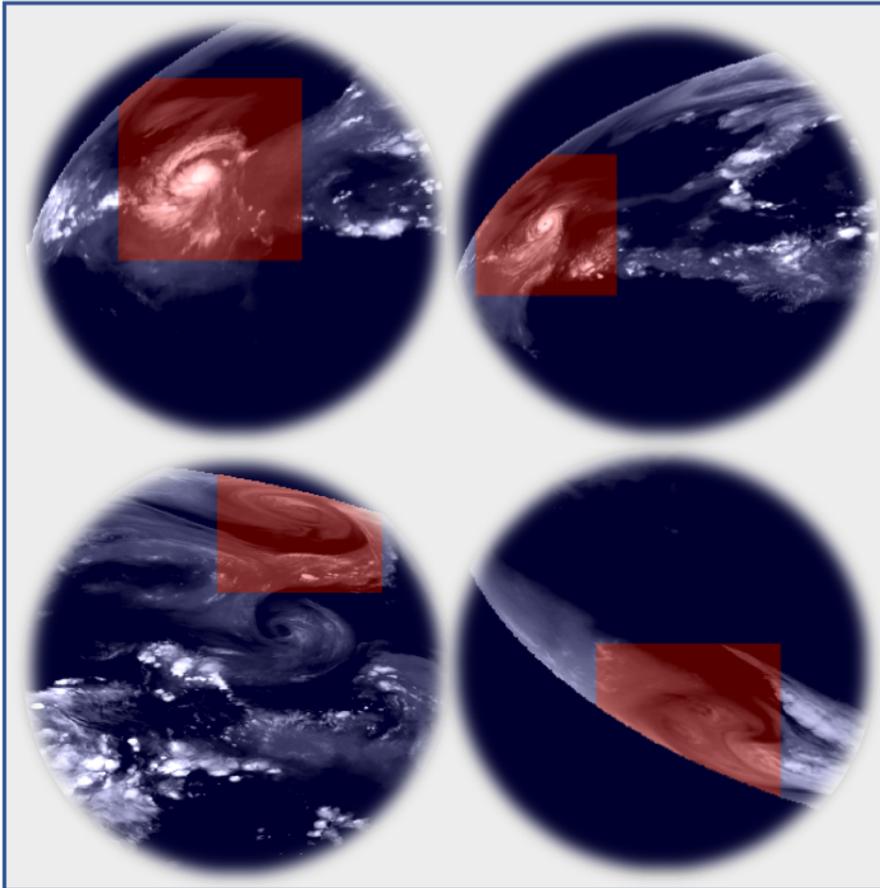

Figure 1. Four different examples of cyclone appearance in the GOES 13 water vapor are highlighted in a red ROI. The top two images are Hurricane Harvey on August 19 (right) and August 21 (left) 2017. Lower two images of extratropical cyclones on August 19 (right) and August 20 (left) 2017 in the northern and southern hemisphere.

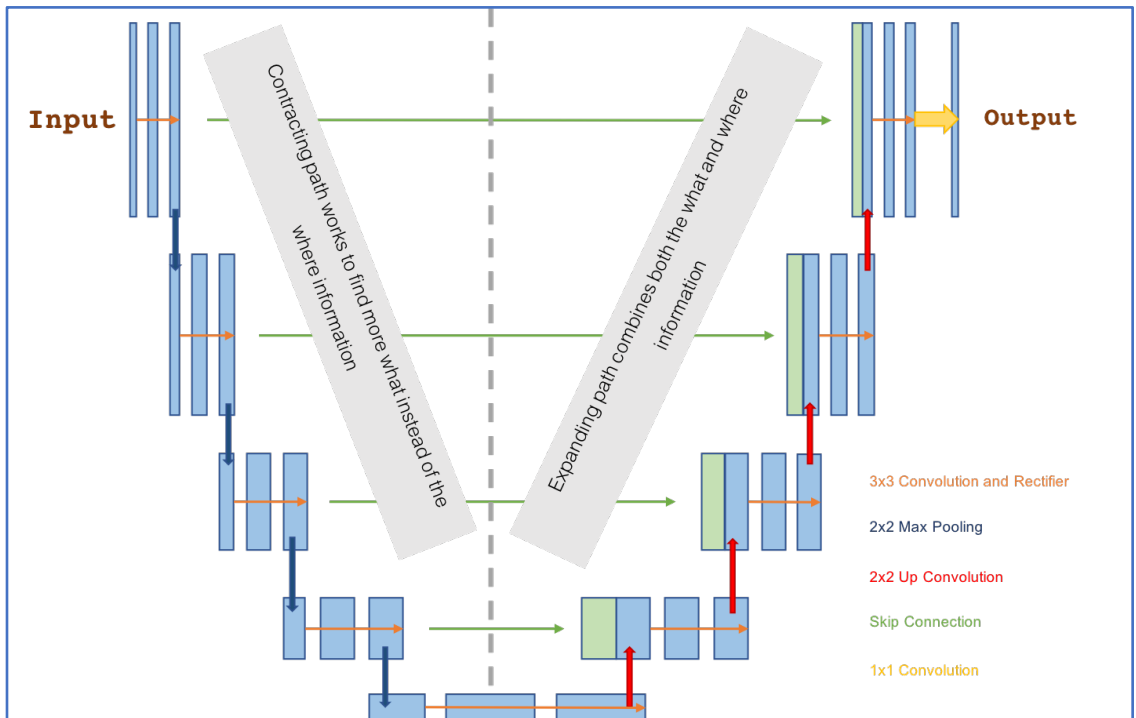

Figure 2. The structure of the U-Net model for all four models. Depth of the model, or how many layers convolution and max pooling for the model, is dynamic and several depths were experimented with to obtain the best architecture for each model. Orange arrows indicate each convolution, dark blue represents the max-pooling layer, green arrows indicate the skip connections and red arrows indicate the up convolutions.

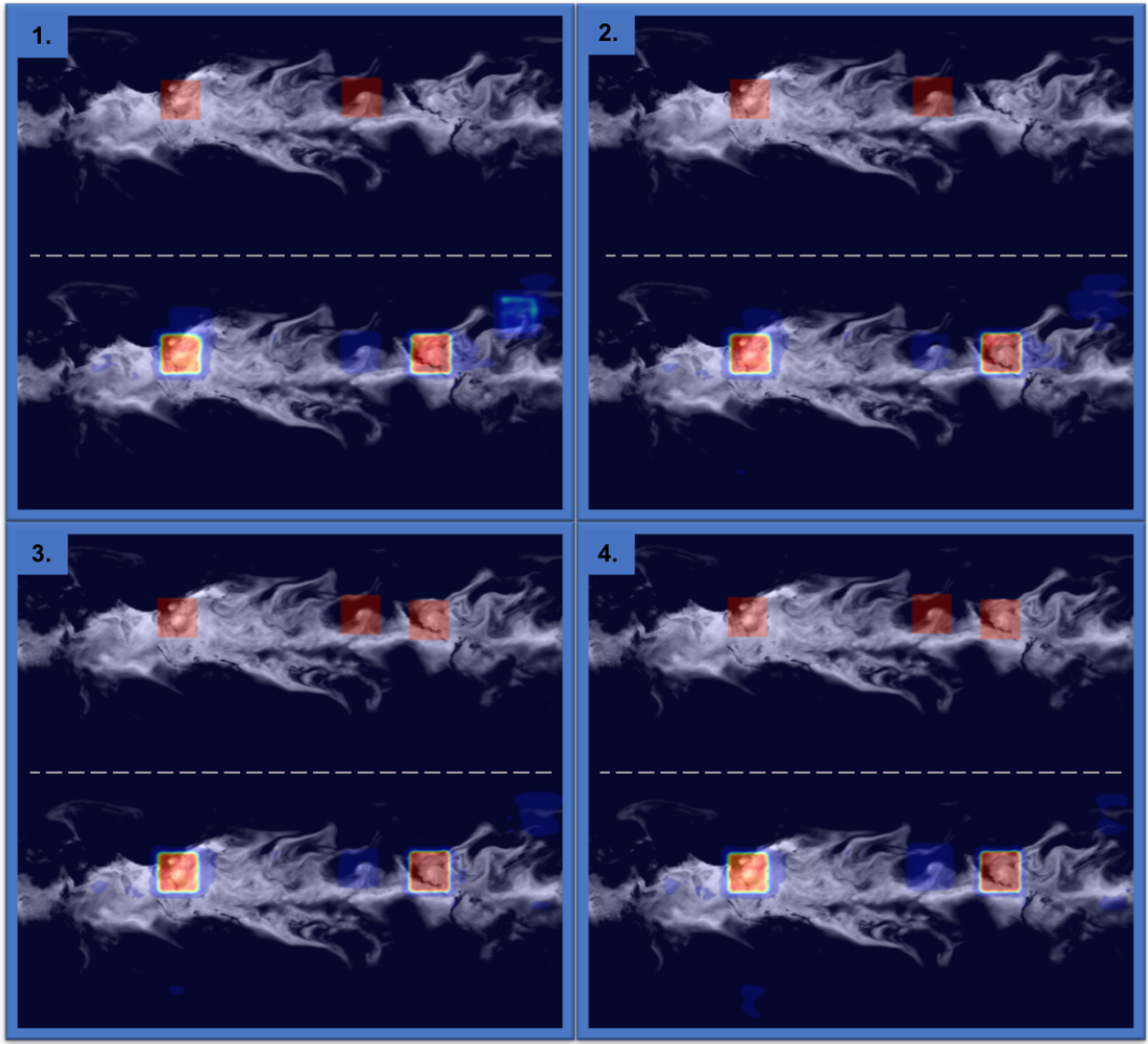

Figure 3. GFS total precipitable water image on August 23rd, 2017. Time progresses in three-hour increments from panel 1 at 12 UTC to panel 4 at 21 UTC. Above the dotted line are the IBTrACS-labeled cyclone ROI used as "truth" with solid red boxes and directly below are the U-Net labeled ROI. In the U-Net results, ROI cyclones are color indicated as a higher "yes-cyclone" likelihood (red) or "not as likely a cyclone" (blue). The background without shading has no cyclone likelihood.

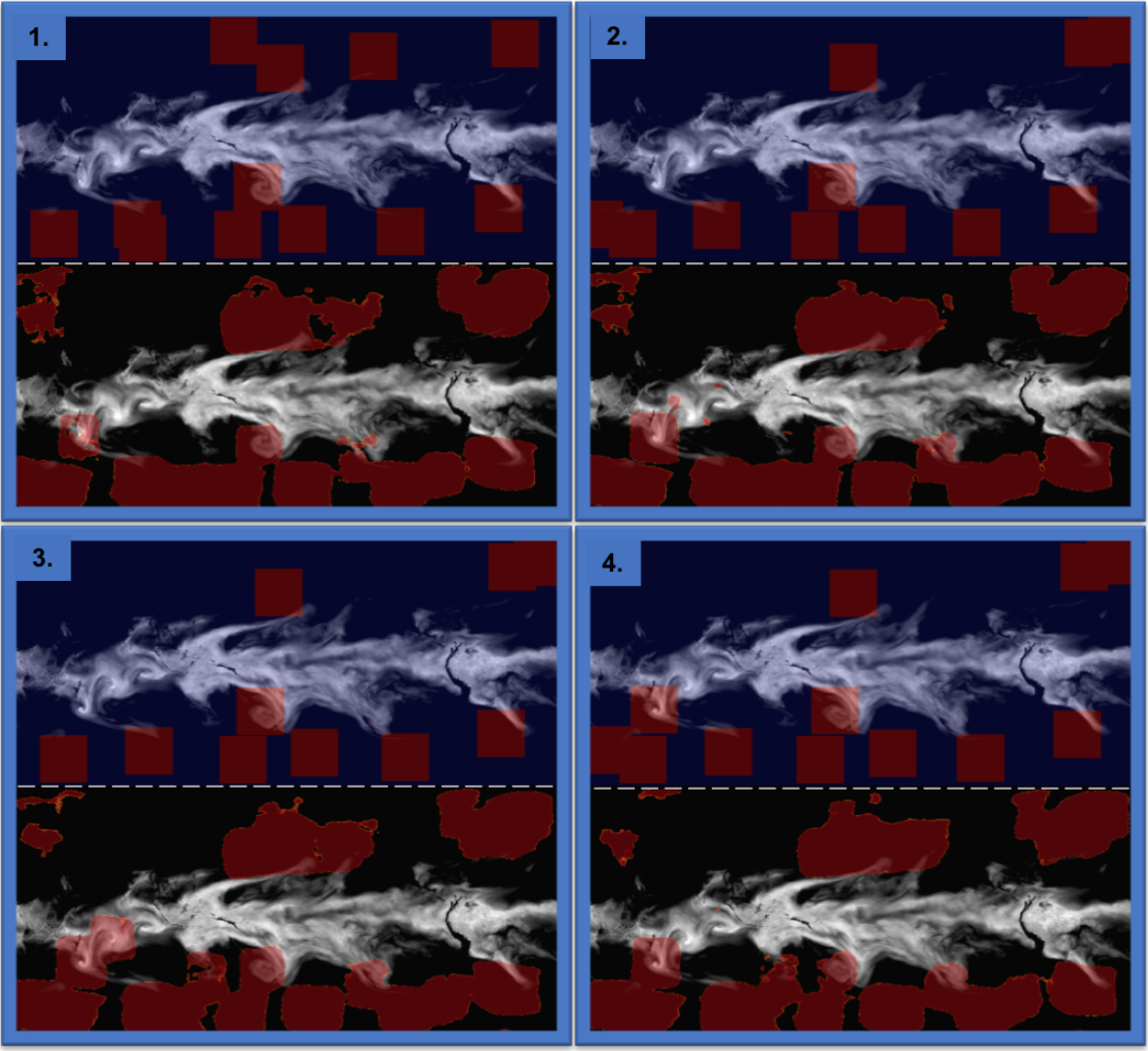

Figure 4. GFS total precipitable water image on March 10th, 2017. Time progresses in three-hour increments from panel 1 at 15 UTC to panel 4 at 0 UTC on March 11th. Above the dotted line are the heuristic-labeled cyclone ROI used as "truth" in labeled in red boxes and below are the U-Net labeled ROI. The U-Net segmentation shown here has a confidence threshold of 70%, meaning that all red segmented regions have a value of at least 0.7 and indicate a higher confidence of a "yes-cyclone" event.

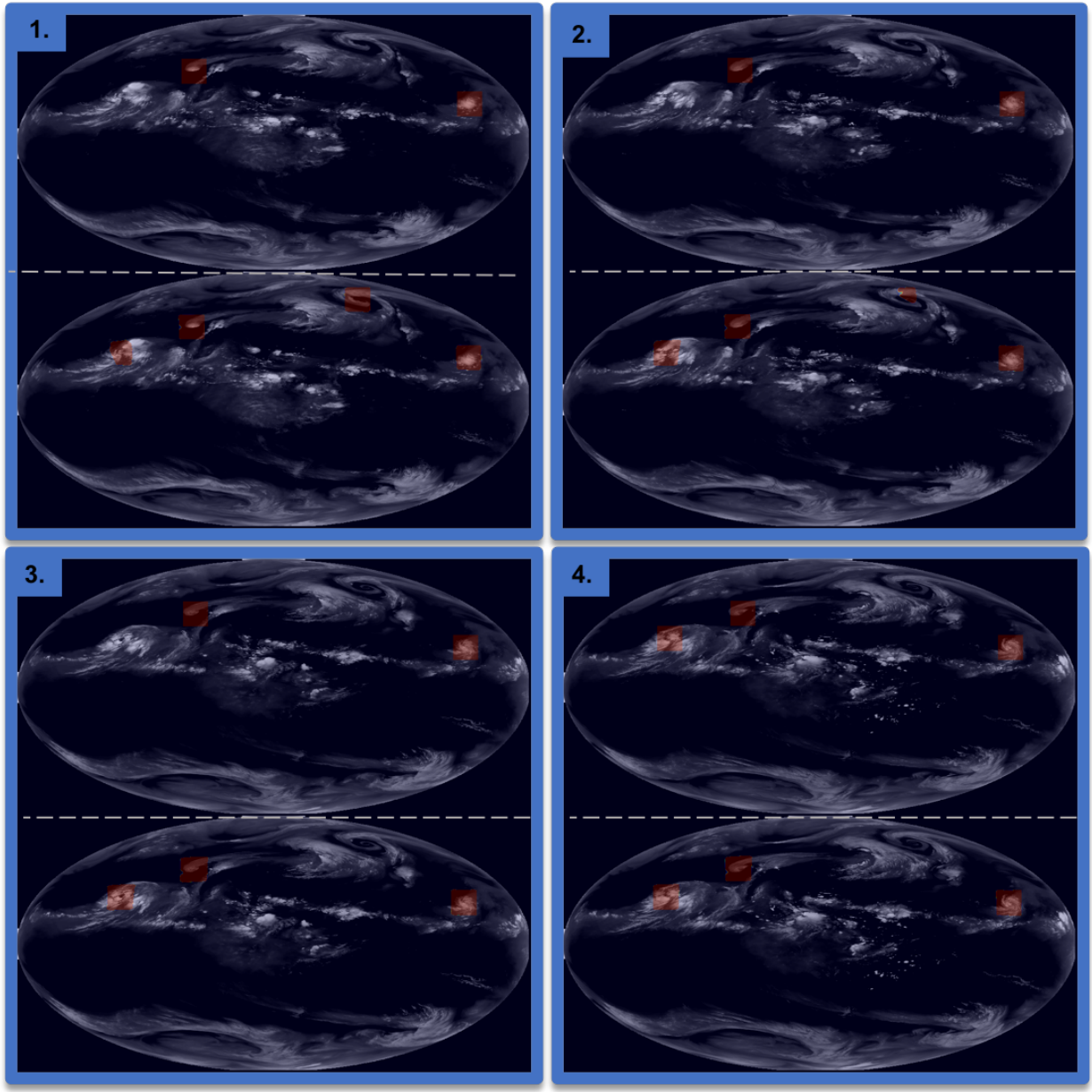

Figure 5. GOES-13 image on August 30th, 2017. Time progresses in three-hour increments from panel 1 at 9 UTC to panel 4 at 18 UTC. Above the dotted line are the IBTrACS-labeled cyclone ROI used as "truth" and below are the U-Net labeled ROI. The U-Net segmentation shown here has a confidence threshold of 70%, meaning that all red segmented regions have a value of at least 0.7 and indicate a higher confidence of a "yes-cyclone" event.

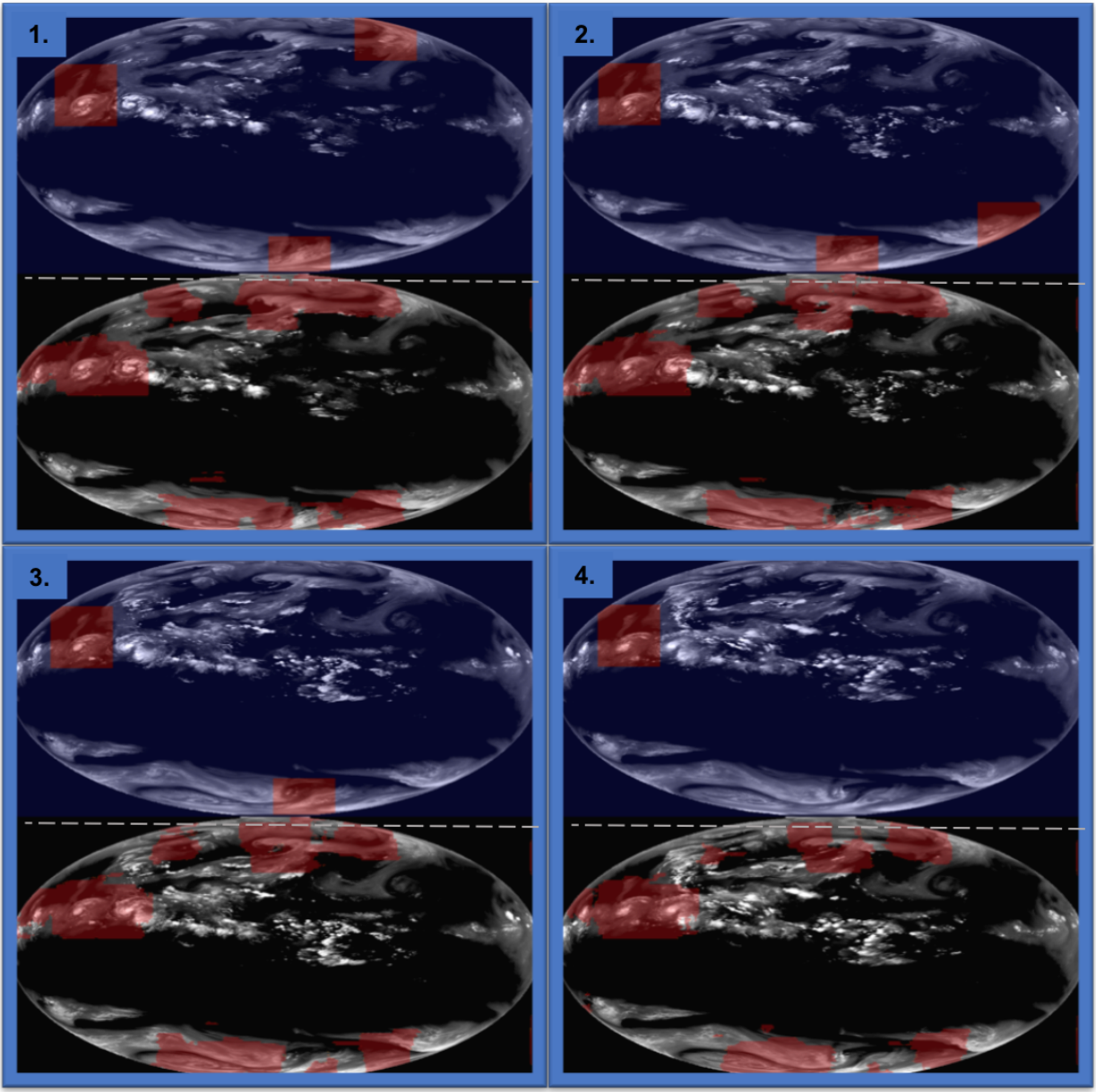

Figure 6. GOES-13 image on July 24th, 2017. Time progresses in three-hour increments from panel 1 at 15 UTC to panel 4 at 0 UTC on July 25th. Above the dotted line are the heuristic-labeled cyclone ROI used as "truth" in red boxes and below are the U-Net labeled ROI in red sections. The U-Net segmentation shown here has a confidence threshold of 70%, meaning that all red segmented regions have a value of at least 0.7 and indicate a higher confidence of a "yes-cyclone" event.

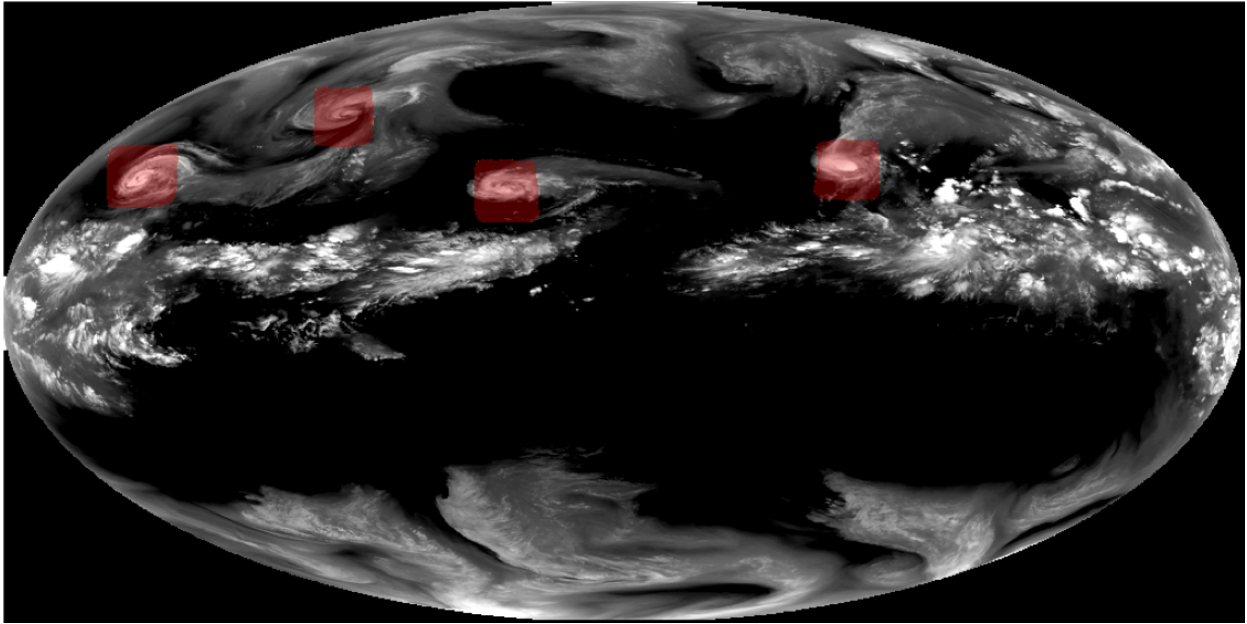

Figure 7. The output from the U-Net trained with IBTrACS labels on GOES water vapor brightness temperature satellite imagery. The regions in red are U-Net identified ROI from September 5th, 2015 at 00UTC

|  | CPU | GPU |
|---|---|---|
| **Hardware** | Two 10 core Haswell Intel | 8 Tesla (P100) |
| **Training (per epoch)** | 11.5 hours | 3 minutes |
| **Complete training (~ 70 epochs)** | ~ 5 weeks | ~ 3 hours |
| **Inference for single input** | 1 second | 40 milliseconds |

Table 1. Comparison between an IBTrACS U-Net model training and inference times on a CPU system compared to running on GPU.

| | | | | | | | | | | | | Inference | Inference |
|---|---|---|---|---|---|---|---|---|---|---|---|---|---|
| | | | | | | | | | | | | Time per | Time per |
| Model | Model | Input Image Size | Training Size | Validation Size | Batch Size | ROI Pixel Size | Dropout or Noise | Loss Function | U-Net Model Depth | Epochs to Convergence | Training Time (seconds) | Single Run (seconds) | Month of Runs (seconds) |
| Labels | Input | | | | | | | | | | | | |
| IBTrACS | GFS | 720x361 | 8622 | 3020 | 256 | 25 | noise 0.2 | Focal | 6 | 37 | 2130 | 0.03 | 8.16 |
| Heuristic | GFS | 720x361 | 15574 | 2902 | 1520 | 30 | dropout 0.1 | Dice | 5 | 200 | 2200 | 0.03 | 6.48 |
| IBTrACS | GOES | 1024X560 | 5638 | 2214 | 128 | 25 | dropout 0.2 | BCE | 5 | 70 | 3540 | 0.15 | 36 |
| Heuristic | GOES | 1024X560 | 25288 | 2735 | 720 | 60 | dropout 0.1 | Tversky | 4 | 150 | 4650 | 0.06 | 14.4 |

**U-Net Model Architectures**

Table 2. A summary of the four selected U-Net models with the specifics of each model. Details include: "truth" label source, input pixel size, number of input images for training and validation, number of images per batch, model activation and loss function, number of epochs to convergence, time for training, and time for inference both on a single input and for a month of inputs.

| | | | | | | | | |
|---|---|---|---|---|---|---|---|---|
| **U-Net Model Results** | | | | | | | | |
| Model Labels | Model Input | Accuracy | Loss Score | Dice Coefficient | Tversky Coefficient | Optimizer | Dropout or Noise | Batch Normalization |
| IBTrACS | GFS | 0.991 | 0.237 | 0.763 | 0.750 | rms 0.00008 | noise 0.2 | yes |
| Heuristic | GFS | 0.807 | 0.351 | 0.58 | 0.649 | rms 0.00001 | dropout 0.1 | yes |
| IBTrACS | GOES | 0.996 | 0.311 | 0.689 | 0.680 | rms 0.0001 | dropout 0.2 | yes |
| Heuristic | GOES | 0.901 | 0.442 | 0.511 | 0.558 | rms 0.00001 | dropout 0.1 | yes |

Table 3. Results from the best four U-Net models with either IBTrACS or Heuristic "truth" labels for both GFS and GOES image inputs. Model performance is measured by Dice and Tversky coefficients as well as accuracy.